\documentclass[twocolumn,english]{IEEEtran}
\usepackage[colorlinks,citecolor=blue,urlcolor=blue,bookmarks=false,hypertexnames=true]{hyperref} 
\usepackage{dblfloatfix}
\usepackage{color}
\usepackage{array}
\usepackage{units}
\usepackage{textcomp}
\usepackage{amsmath}
\usepackage{graphicx}
\usepackage{amssymb}
\usepackage{times,amsmath,epsfig}
\usepackage{graphicx}
\usepackage{gensymb}
\usepackage{amsmath}
\usepackage{subfigure}
\usepackage{psfrag}
\usepackage{epsfig}
\usepackage{cite}
\usepackage{multirow}
\usepackage{multicol}
\usepackage{babel}
\usepackage{indentfirst, float}
\graphicspath{ {images/} }
\usepackage{caption}
\usepackage{picinpar}
\usepackage{url}
\usepackage{flushend}
\usepackage{colortbl}
\usepackage{soul}
\usepackage{pifont}
\usepackage{alltt}

\title{Blockchain-based Immutable Evidence and Decentralized Loss Adjustment for Autonomous Vehicle Accidents in Insurance}

\author{ \href{https://orcid.org/0000-0003-0276-9289}{\includegraphics[scale=0.06]{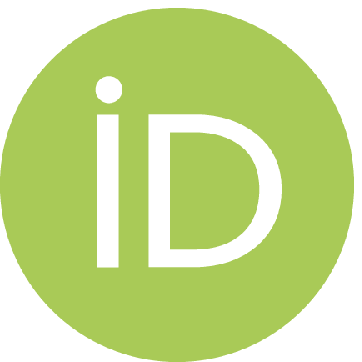}\hspace{1mm}Mehmet Parlak}
\thanks{The authors are at the Department of Electrical and Electronics Engineering, Ozyegin University, Istanbul (e-mail: mehmet.parlak@ozyegin.edu.tr)} \\
	Department of Electrical and Electronics \\
	Ozyegin University\\
	Cekmekoy, Istanbul 34794 \\
	\texttt{mehmet.parlak@ozyegin.edu.tr} \\
 \vspace{-5mm}
}

\hypersetup{
pdftitle={A template for the arxiv style},
pdfsubject={q-bio.NC, q-bio.QM},
pdfauthor={Mehmet Parlak},
pdfkeywords={First keyword, Second keyword, More},
}

\begin{document}
\maketitle
\begin{abstract}
In case of an accident between two autonomous vehicles equipped with emerging technologies, how do we apportion liability among the various players? A special liability regime has not even yet been established for damages that may arise due to the accidents of autonomous vehicles. Would the immutable, time-stamped sensor records of vehicles on distributed ledger help define the intertwined relations of liability subjects right through the accident? What if the synthetic media created through deepfake gets involved in the insurance claims? While integrating AI-powered anomaly or deepfake detection into automated insurance claims processing helps to prevent insurance fraud, it is only a matter of time before deepfake becomes nearly undetectable even to elaborate forensic tools. This paper proposes a blockchain-based insurtech decentralized application to check the authenticity and provenance of the accident footage and also to decentralize the loss-adjusting process through a hybrid of decentralized and centralized databases using smart contracts.

\end{abstract}


\section{Introduction}
\begin{figure}[htbp!]
	\centering
	\noindent \includegraphics[width=0.47 \textwidth]{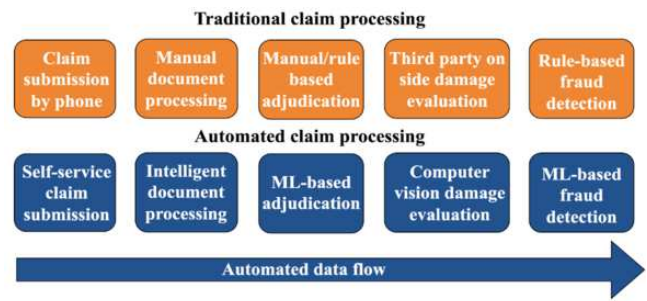}\\
	\caption {Comparing traditional and automated insurance claim processing}
	\label{fig:intro}
\end{figure}

Artificial intelligence (AI) is transforming the insurance industry through machine learning-based file, image, and video analysis, as shown in Fig. \ref{fig:intro}, wherein insurers can now automatically process and settle claims at a fast pace and provide estimates with greater accuracy. A shorter claim settlement cycle time results in higher customer satisfaction in return \cite{Claims20}. Machine learning (ML) algorithms programmed to curate relevant data from the claims-related photos and videos can accurately gauge the extent of damage and automate the claims assessment process further \cite{Patil21},\cite{Cha17}, \cite{Liu18}, \cite{Singh19}. 

However, deepfake threatens insurance claims automation through AI-generated synthetic images and videos without forensic traces typically found in edited media \cite{Collins19}. Researchers have concluded it is only a matter of time before deepfake becomes nearly undetectable to the human eye and even to elaborate forensic tools \cite{Mangaokar21}. In a recent survey, over 80\% of insurance professionals indicated that altered or tampered digital media that falsely inflate insurance transactions, such as claims, are the top concern among the various types of media-related fraud, which is an explicit acknowledgment of the fraudulent losses in a market already suffering from over \$80B in annual fraud just in the US \cite{Attestiv21}. 

The pace of insurance claims automation exceeding the pace of automated fraud prevention creates new risks of fraud. As a result, implementing automated fraud prevention tech. is quickly becoming essential for insurance companies trying to protect their business metrics. The Coalition for Content Provenance and Authenticity (C2PA) - a joint development foundation project - and the Content Authenticity Initiative - a cross-industry initiative - have released industry standards for content tracking, provenance, and authenticity of digital media files  \cite{C2PA}. AI-generated images and videos call for protective action by insurers to defend against this burgeoning fraud. Insurtech ecosystem is exploring diverse and scalable solutions to this challenge, such as integrating AI-powered fraud detection into automated claims processing, which is called “in-line detection,” and also exploring other technologies like blockchain to establish the provenance of digital media, which is called “in-line prevention” \cite{Attestiv22},\cite{Aslam22},\cite{Hasan19}.

In-line detection is the only defense when the digital media is not captured by a trusted application or person. However, it would require a reasonable amount of time and processing power due to extensive AI analysis, given constant improvements in deepfake technology. On the other hand, the prevention only applies at the point of capture, making it useless when capture software is unavailable or not used. Therefore, a hybrid arrangement is recommended, starting with an application that captures the claim-related media, and authenticates them at the point of capture to remove the threat of deepfake while keeping a counter-deepfake tool for in-line detection as two-factor authentication, if needed.

The rest of the paper is organized as follows: Section II provides background about blockchain technology, hashing, smart contracts, and the Ethereum protocol. In Section III, Tamper-Proof, a decentralized application (dApp) for establishing the authenticity of digital content and decentralizing loss-adjusting, is illustrated to address the challenges in automated insurance claim processing. Section IV provides a futuristic overview of the decentralized, autonomous, and connected insurance market, and finally, section V summarizes the paper. 

\section{Background}
\label{sec:background}

\subsection{Hash}
Hash is a cryptographic function that turns any input file into a string of bytes with a fixed length and structure (hash value). The input can not be obtained back from its hash value because two different inputs with the same hash value cannot be found with the available computation to crack strong collision-free hashing algorithms as of now. In other words, if a single bit in the input message is changed, the hash value will be completely different. Since no two inputs will theoretically have the same hash value \cite{Maurer75}, the hashing methodology can be used to uniquely identify a specific input content storing its unique “digital fingerprint” and prove that a piece of text, content, or file has not been altered over time. 

\subsection{Blockchain}
\begin{figure}[htbp!]
	\centering
	\noindent \includegraphics[width=0.47 \textwidth]{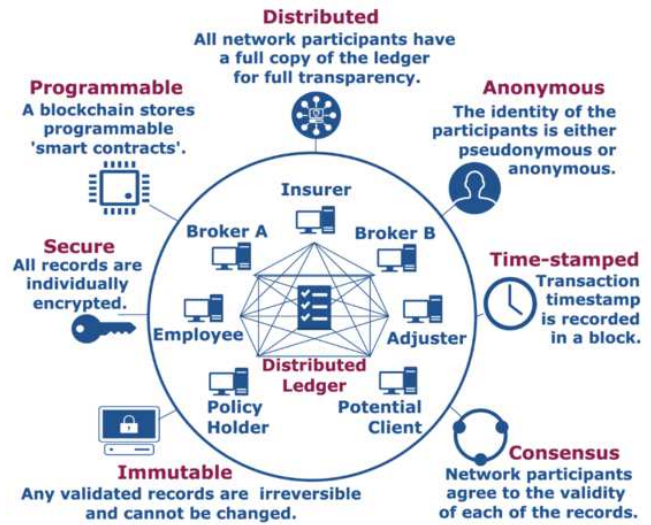}\\
	\caption {Benefits of blockchain in the insurance industry}
	\label{fig:benefits}
\end{figure}

The “chain” of transactions kept in a ledger is known as the blockchain. The transactions are verified and maintained by all of the computers participating in the network and each transaction is essentially a public timestamp that can contain data. The critical aspects of the timestamped transactions are: 

• decentralized (no entity controls the database of timestamps, and everyone in the network confirms that a timestamp has happened), 
• immutable (once a timestamp has been verified and recorded, no one can undo it), 
• public (all of the timestamps are publicly visible, although some aspects of the data are encrypted), and 
• programmable (triggering some sort of action based on the details of a “smart contract” embedded in a timestamp).
Each of the timestamps contains a packet of data that can hold many things such as details about a financial transaction among the parties, or a hashed version of almost any document. While cryptographic algorithms and hashing are not unique to the blockchain, these particular features of the technology are made more powerful once combined with the immutable nature of blockchain technology. Fig. \ref{fig:benefits} summarizes the various benefits that insurtech players can derive by applying blockchain technology to the insurance processes, as blockchain has brought about transparency and transformation in client onboarding, underwriting, and claims processing.

\subsection{Smart Contracts}

A smart contract (SC) is a computerized transaction protocol that executes the terms of a contract on a distributed ledger such as a blockchain. It is not necessarily a contract in a legal sense; however, it works the same way IT systems execute the contract terms today. A developer initially programs the SC to facilitate, verify, and enforce/execute the negotiation and agreement automatically among multiple untrustworthy parties via decentralized applications (dApps) so that all participants can be immediately certain of the outcome and minimize exceptions, both malicious and accidental without any trusted intermediary's involvement or time loss. dApps are blockchain-powered software, whereas smart contracts work as application programming interface (API) connectors that connect dApps with blockchain.
To establish the terms, participants must determine how transactions and their data are represented on the blockchain, agree on the “if/when...then…” statements that govern those transactions, explore all possible exceptions, and define a framework for resolving disputes. There can be as many stipulations as needed to satisfy the participants that the task will be completed satisfactorily through the common contractual conditions (such as payment terms, liens, confidentiality, and even enforcement) through a dApp tailored to the sectors such as insurance, finance, logistics, transport, energy production, health systems, education, and public service management. When a network of computers executes the actions and predetermined conditions have been met and verified, the transaction cannot be changed, and only parties who have been granted permission can see the results.

\subsection{Ethereum v2.0}

\begin{figure}[htbp!]
	\centering
	\noindent \includegraphics[width=0.47 \textwidth]{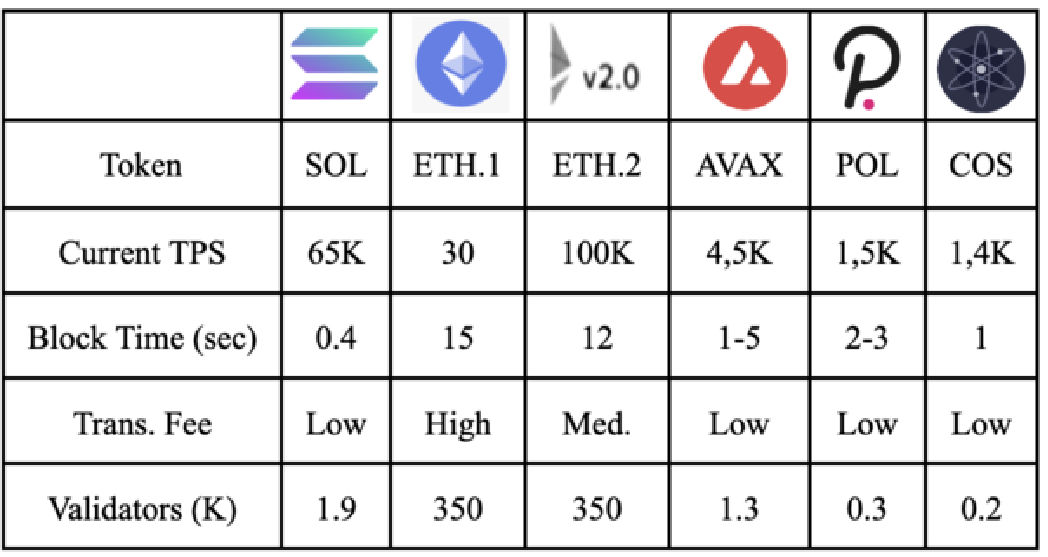}\\
	\caption {Performance Comparison of Layer-1 Protocols}
	\label{fig:compare}
\end{figure}

Through a cryptographic proof mechanism called the proof of work (PoW), the Ethereum 1.0 platform used to execute a probabilistic consensus between Ethereum clients connected in a mainnet, reaching an agreement on a transaction’s validity and providing the platform with security and decentralization. However, highly variable fees and slow confirmation times due to a growing user base, as shown in Fig. \ref{fig:compare}, have caused congestion and been a bottleneck for the widespread adoption of smart contracts on Ethereum.

\begin{figure}[htbp!]
	\centering
	\noindent \includegraphics[width=0.47 \textwidth]{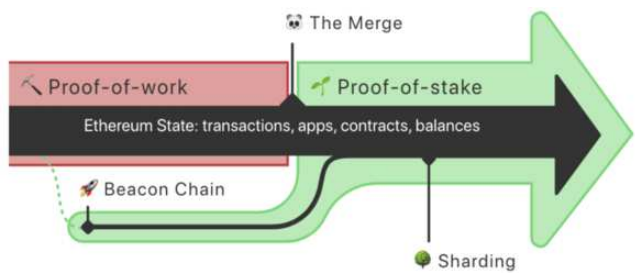}\\
	\caption {Ethereum mainnet merging with the proof-of-stake consensus layer, the Beacon Chain}
	\label{fig:mainnet}
\end{figure}

As shown in Fig. \ref{fig:mainnet}, the current Ethereum mainnet merging with the proof of stake consensus layer, the Beacon Chain ends the proof of work for Ethereum and enables the public ledger to operate more efficiently. The Beacon Chain, developed in December 2020,  has since existed as a separate blockchain to Mainnet, running in parallel but it has not been processing Mainnet transactions \cite{Ethereum}. Instead, it has been reaching a consensus on its own state by agreeing on active validators and their account balances. The “Merge” represents the official switch to using the Beacon Chain as the engine of block production and the consensus engine for all network data, including execution layer transactions and account balances. Mining will no longer be the means of producing valid blocks. Instead, the proof-of-stake validators assume this role and will be responsible for processing the validity of all transactions and proposing blocks. 

Ethereum 2.0 provides future scaling upgrades including 64 sharding chains, extending the network with more chains, which run in parallel. Sharding is a way of spreading out the computing and storage workload from a network so that each node doesn’t have to process the entire network’s transactional load. Each Ethereum node will only have to run one of the shards, which means only a small amount of data has to be computed without any need for powerful hardware. This makes it easier for users to operate nodes that secure the Ethereum network. Sharding will improve scalability by enabling further layer-2 scaling, in particular Roll-ups. Roll-up transactions are executed on a separate chain, then they are bundled together and submitted to the main Ethereum chain. This essentially means, only some of the data of the roll-up transactions has to fit into the blocks of the main chain. In other words, the Beacon Chain can be understood as the coordination layer and the shards are the data layer. 

The new mechanism will allow Ethereum to scale from 30 to 100,000 transactions per second (TPS), compared to the other competitors shown in Fig. \ref{fig:compare}. Through the “Merge” the need for energy-intensive mining is eliminated, and Ethereum's energy consumption is reduced by ~99.95\%. Furthermore, the ETH proof-of-stake mechanism brings further scalability (higher throughput), sustainability (lower fees, lower energy consumption), and security (large pool of ETH validators). 

In addition, the shift to Ethereum 2.0 encourages more users to participate in the network as validators, leading to a more decentralized and secure system. With the rollout of Ethereum 2.0, the platform is well-positioned to attract more developers and users to its ecosystem, providing new opportunities for innovation and growth in the blockchain industry.

\vspace{3mm}

\section{Tamper-proof}
\label{sec:tamper}

\subsection{Emerging Technologies around Tamper-Proof}

 In an ecosystem where both vehicles and smart contracts work autonomously, it is inevitable that determining the source and also the compensation of the damage shall be autonomous and decentralized, with the inclusion of new technologies. Via sensors and emerging technologies installed on the vehicles, AV manufacturers and software providers not only ensure the autonomous movement of the vehicle but also ensure that AV collects sufficient data for liability association before and during the event of an accident. 

When the accident occurs, the Tamper-Proof dApp installed in the AV utilizes the internet of things (IoT) and artificial intelligence (AI) technologies to instantaneously communicate with the surrounding AVs to ensure that captured video footage (before and after the crash) of all the neighboring vehicles are collected rather than getting erased per privacy and data regulations dictate.

Furthermore, storing such data in a centralized database requires that the database is adequately guarded and the data itself will not be deleted or tampered with after being written. Therefore, a solution that is not controlled by one single entity but rather distributed and decentralized across many neutral parties is needed to build a certain level of trust. 

Tamper-Proof is an insurtech dApp that establishes a set of authentic tamper-proof records, and decentralized loss-adjusting for insurance claims processing and resolution. It benefits from the Ethereum network to sign the issuing device’s keys, manage the flow of the metadata inputted by the user, and receive a consensus timestamp, as shown in Fig. \ref{fig:tamper}. Tamper-Proof, continues to be built on Ethereum 2.0 after the Merge. The nature of the proof of stake consensus will enable the provenance of large amounts of digital media to eliminate insurance fraud and scale automated insurance transactions through SC technology. 

\subsection{Decentralized Tamper-Proof Evidence}

Tamper-Proof dApp provides valuable forensic evidence such as the time of the incident, and the location of the vehicles, together with the uploads of footage video, photo, and audio files at scale. Typically, all this data would be stored in a centralized database, increasing the probability of data loss or manipulation. 

\begin{figure*}[b!]
	\centering
	\noindent \includegraphics[width=1 \textwidth]{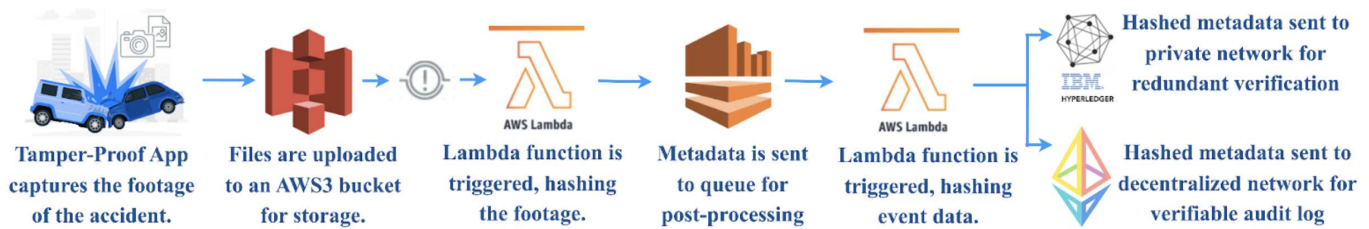}\\
	\caption {Tamper-Proof information flow to capture and secure the provenance of the accident footage}
	\label{fig:tamper}
\end{figure*}

\begin{figure}[t!]
	\centering
	\noindent \includegraphics[width=0.47 \textwidth]{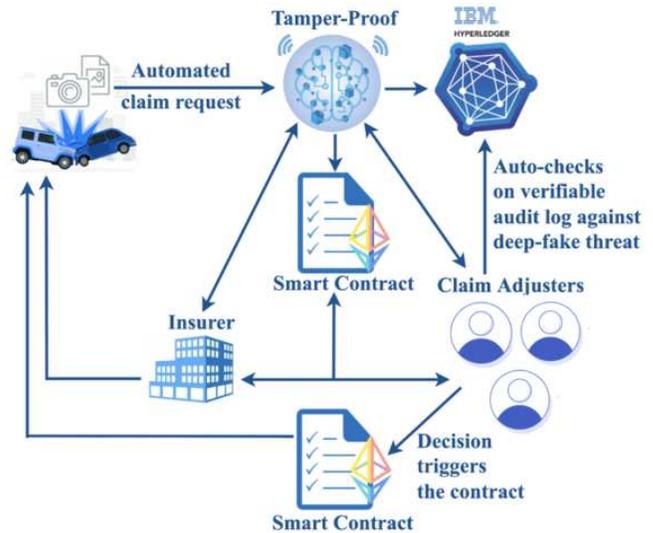}\\
	\caption {Accident footage authentication and decentralized loss adjusting via automated claims on Tamper-Proof}
	\label{fig:scene}
\end{figure}

As shown in Fig. \ref{fig:scene}, the stakeholders involved in the claim processing phase are clients, claim loss adjusters, and insurance companies. The loss adjusters are employed by the insurance company but they are supposed to remain independent. In the case of a fully autonomous vehicle (AV) accident, the AV acts as the “client” and makes a claim automatically attaching the footage of the accident and claim-related media as supporting evidence into the cloud. After the claim has been made, the loss adjusters get a notification. They start the verification and investigation process to determine the validity of the claim and the amount of loss or damages covered by the insurance policy. They also perform in-depth verification and find out if the evidence submitted by the client is valid or not, where Tamper-Proof provides a significant value proposition.

Through the  Tamper-Proof, the incident video footage is automatically sent to Amazon Simple Storage Service (AWS S3 bucket), which is an object storage service built to store and retrieve data from anywhere, as shown in Fig. \ref{fig:tamper}. For Tamper-Proof to enhance the trustworthiness of the video footage, a hash of it is created via a triggered AWS Lambda function. AWS Lambda is used to create backend application services triggered on demand using the Lambda application programming interface (API). Next, the output hash is sent to a Kinesis message queue, a fully manageable message queuing service for real-time streaming data. This hash flows through Amazon Kinesis via another triggered AWS Lambda function, and then it is simultaneously written both into a private ledger, IBM Blockchain Platform powered by Hyperledger Fabric, and Ethereum Platform. Tamper-Proof uses the IBM Blockchain Platform as its internal repository, preserving additional metadata to augment the auditable hash data sent to the Ethereum platform. With two sets of records, Tamper-Proof stakeholders can verify that the image and its associated metadata were not altered or tampered with. Any third party can verify the information on the Ethereum public ledger to match the image and its accompanying data stored on the private ledger, if necessary and if permission is given.

If the footage of the incident is found to be authentic and accurate, adjusters validate the initial claim request and add related details to the blockchain. The adjusters will also identify any witnesses around the incident, access their verified footage, and reach the parties for extra information if needed. Once the investigation is complete, the adjusters go through the policy carefully to determine what is and isn’t covered under the policy, and inform the client of any applicable deductibles that may apply to the case.

\subsection{Decentralized Loss Adjusting}

Insurance claims are gathered and investigated by loss adjusters on behalf of the insurer. Tamper-Proof leverages the capabilities of blockchain technology to create an alignment of economic incentives, which in turn results in the predictability of the behavior of loss adjusters. Becoming a loss adjuster in Tamper-Proof requires a certificate and investment of capital, called “staking.” By staking the tokens of Tamper-Proof, anyone certified and accredited in loss-adjusting can become a part of the pool of adjusters and is eligible to participate as an independent adjuster when called upon to do so. Once an adjuster is called upon to investigate a claim, the adjuster first considers the authenticity of the evidence presented after inspecting the damage. Then, the adjuster investigates who is liable for the damage, verifies if the policyholder is liable to get the claim amount, and determines a fair amount for compensation in favor of either counterparty.

Via Tamper-Proof, insurance claim investigations are assigned to not one but several loss adjusters in the loss-adjusters pool for further decentralization, which creates an attractive online insurance claim investigation model; decentralized, fast, and automated. The consensus among the selected loss adjusters determines the outcome of the insurance claim, where the economic interests of the loss adjusters are aligned to make coherent and correct decisions. 

\begin{figure}[htbp!]
	\centering
	\noindent \includegraphics[width=0.47 \textwidth]{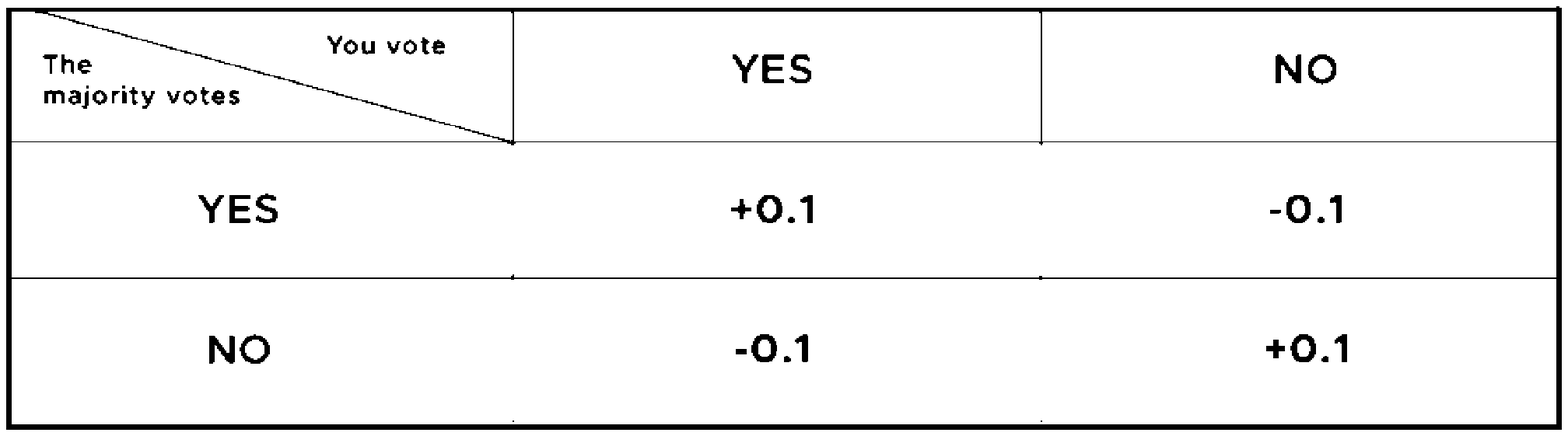}\\
	\caption {Payoff table for a basic Schelling game
}
	\label{fig:game}
\end{figure}

The consensus and redistribution mechanism in Tamper-Proof is inspired by Thomas Schelling, a game theorist. Schelling invented the SchellingCoin concept, where jurors gain or lose tokens depending on whether their vote was consistent with the other jurors, as shown in Fig. \ref{fig:game} \cite{Schelling58}. Schelling games are the basis for many blockchain components that are used to import information onto a blockchain platform, although they are not checked as part of the consensus protocol of that blockchain, such as "oracles." Applying the Schelling Point concept to the claim adjusting, adjusters are incentivized to decide honestly because, after a dispute is over and after Tamper-Proof has reached a decision on the dispute, tokens are unfrozen and redistributed among jurors with coherent decisions. The adjusters whose decisions are not coherent with the majority of the adjusters will not receive their arbitration fees and lose some Tamper-Proof tokens.

After loss adjusters/auditors validate the request, the insurance company sends the claim breakdown to the consumer, as shown in Fig. \ref{fig:game}, Once the consumer receives the breakdown of the amount, the claim money is automatically transferred to them via automated smart contracts. Blockchain can reduce the turnaround time by simplifying, and speeding up the entire insurance claim process.

\subsection{Decentralized, Autonomous, Connected Auto-insurance}

In the next-generation AV insurance ecosystem, the convergence of new players and disruptive technologies such as smart contracts and oracles empowered through distributed ledger technologies (DLT) such as blockchain, as well as IoT, and AI will disrupt the industry taking an active role in determining the source of the damage. While blockchain technology provides new business models through its transparent, secure, and robust consensus and accountability mechanisms, IoT and AI provide AV-driving dynamic risk pricing and cognitive, agile, scalable claim processing and underwriting, respectively.

While the new technologies help determine the source of the damage, they might also cause damage with various security problems they bring in. The increased connectivity in AVs may also expose the vehicle to attacks from interactive entities, motivating potentially responsible entities to decline liability. Therefore, AVs envisage a much more complicated liability model by assigning responsibility to new technology players. While the factors affecting the accident may differ for each Level-5 vehicle, with the sensors and internet communication devices/technologies installed on the AVs, AV manufacturers and software providers will ensure that the AVs collect sufficient footage data as evidence and for liability association in the event of an incident. The autonomous, decentralized, time-stamped, and tamper-proof recording of all interactions between all players in the AV ecosystem on a distributed ledger will help to determine the contractual and "non-contractual" legal liability, mainly if AVs cause harm to third parties participating in the traffic.

This technological revolution also presents a considerable threat and a necessity for traditional auto-insurers to obtain and control the data generated by AVs and the communications and software systems that support them. Insurers will need to effectively identify, map out and collaborate with the potential partners in the AV ecosystem that can collect, organize and analyze the sensor data, such as OEMs, communication and software systems providers, and even governments at multiple levels. 

Auto insurers insuring thousands of small risks have revenues derived primarily from personal automobile policies. With the convergence ecosystem in technology, autonomy, and mobility, they need to transform themselves into large commercial insurers writing policies on a small number of substantial risks such as cyber security, product liability for sensors and software, and infrastructure problems \cite{Accenture}. As AVs shift the industry focus from personal ownership and liability to commercial and product liability, the most significant payers of new premiums will become OEMs, technology giants, and the government. 

As documented in the literature [17-23], in Table II, there have been numerous works proposing the joint use of blockchain and smart contracts in the vehicular ecosystem and insurance. The blockchain-based smart contracts empowering the next-generation insurtech dApps in the AV ecosystem will disrupt the insurance industry and play a fundamental role in formalizing the contractual and tort liability between the players of the AV ecosystem; AV manufacturer, software provider, operator, users, and insurer.

\begin{table*}[t!]
\caption{Related Works of Emerging Tech-Based Insurance Services}
\label{Perfcomparison}
\renewcommand{\arraystretch}{2.5}
\begin{centering}
\begin{tabular}{|>{\centering}m{65pt}|>
{\centering}m{16pt}|>
{\centering}m{107pt}|>
{\centering}m{270pt}|}
\hline 
Dorri \textit{et al.} \cite{Dorri17} & 2017 & Blockchain  & Protects the privacy of users and increases the security of the vehicular ecosystem.
\tabularnewline
\hline 
 Bader \textit{et al.} \cite{Bader18} & 2018 & Blockchain and smart contracts  & Enables cost savings by removing manual inspection of insurance claims.
\tabularnewline
\hline 
Cebe \textit{et al.} \cite{Cebe18} & 2018 & Blockchain  & Stores maintenance history of vehicles and enables trustless post-accident analysis.  \tabularnewline
\hline 
Oham \textit{et al.} \cite{Oham18} & 2018 & Blockchain  & Tracks both sensor data and entity interactions with two-sided verification.
\tabularnewline
\hline 
Lamberti \textit{et al.} \cite{Lamberti18} & 2018 & Smart contracts and sensors  & Provides on-demand coverage. 
\tabularnewline
\hline 
Dhieb \textit{et al.} \cite{Dhieb20} & 2020 & Permissioned Blockchain and AI  & Secures insurance activities, and detects fraudulent claims.
\tabularnewline
\hline 
 Kong \textit{et al.} \cite{Kong21} & 2021 & Blockchain & Achieves the verifiable aggregation and immutable dissemination of records.
\tabularnewline \hline 
 Parlak \textit{et al.} \cite{Parlak23} & 2022 & Blockchain & Immutable Evidence and
Decentralized Loss Adjustment.
\tabularnewline
\hline 
\end{tabular}
\par\end{centering}
\end{table*} 
\section{Conclusion}

AI-generated images and videos call for protective action by insurers to defend against insurance fraud. With the deep learning technologies behind deepfake getting more sophisticated, integrating and utilizing AI-powered deepfake detection in automated insurance claims processing (in-line detection) requires a reasonable amount of time and processing power due to extensive AI analysis. Therefore, utilizing blockchain to establish the authenticity and provenance of the on-site captured, claim-related digital media (in-line prevention) is recommended to remove the threat of deepfake. 

In the first part of this paper, we have illustrated an insurtech dApp that captures and secures the provenance of the accident footage, provides tamper-proof and authentic evidence of the claim-related media for scaling and automating insurance claim processing. In the second part, a decentralized economic incentivized loss adjusting model is illustrated to enhance fair and objective claim adjusting processes. By incorporating blockchain technology, insurers can create an immutable and transparent record of all claim-related media, preventing the possibility of fraud or tampering. .

The decentralized nature of blockchain ensures that there is no single point of failure, making it virtually impossible for bad actors to manipulate the system. In addition, the economic incentivized loss adjusting model provides a fair and objective approach to claim adjusting, rewarding honest policyholders and ensuring that fraudulent claims are caught and penalized. By leveraging the power of blockchain and AI technologies, insurers can create a more secure and efficient claims processing system, ultimately providing better service and protection for their customers. The combination of these technologies has the potential to revolutionize the insurance industry and improve its overall performance, reducing costs and increasing trust in the system.

\bibliography{references}
\bibliographystyle{IEEEtran}
\end{document}